\documentclass[journal=preprint,manuscript=article]{achemso}

\usepackage[version=3]{mhchem} 
\usepackage{amsmath,amssymb}
\usepackage{graphicx}
\usepackage{float}
\usepackage{hyperref} 
\usepackage{xcolor}
\usepackage{ulem}
\author{P.R.A. de Oliveira}
\affiliation{Instituto de Física, Universidade Federal do Rio de Janeiro (UFRJ), 21941-909, Rio de Janeiro, RJ, Brazil}
\alsoaffiliation{Centro Brasileiro de Pesquisas Físicas (CBPF), 22290-180, Rio de Janeiro, RJ, Brazil}
\email{hninofilho@gmail.com}

\author{C. Codeço}
\affiliation{Instituto de Física, Universidade Federal do Rio de Janeiro (UFRJ), 21941-909, Rio de Janeiro, RJ, Brazil}

\author{M.G.Menezes}
\affiliation{Instituto de Física, Universidade Federal do Rio de Janeiro (UFRJ), 21941-909, Rio de Janeiro, RJ, Brazil}

\author{P. Venezuela}
\affiliation{Instituto de Física, Universidade Federal Fluminense (UFF), Campus da Praia Vermelha, Niterói, RJ, 24210-346, Brazil}

\author{F. Stavale}
\affiliation{Centro Brasileiro de Pesquisas Físicas (CBPF), 22290-180, Rio de Janeiro, RJ, Brazil}

\author{J.A.Boscoboinik}
\affiliation{Brookhaven National Laboratory, Upton, NY, 11973, USA}

\title{Co-adsorption mechanism drives CO oxidation on defective ZnS}

\keywords{NAP-XPS,DFT. CO oxidation, ZnS, defects}

\begin{document}

\begin{tocentry}
   \includegraphics{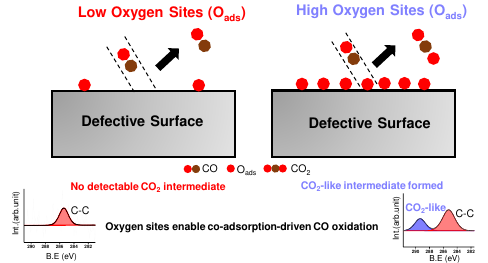}
\end{tocentry}

\begin{abstract}
Reactivity on wide-bandgap semiconductor surfaces relies critically on the generation of active sites. In the case of CO oxidation, however, the mere presence of defects is insufficient to drive reactivity. Here, we investigate CO oxidation on a defective ZnS single-crystal surface by combining near ambient pressure X-ray photoelectron spectroscopy (NAP-XPS) and density functional theory (DFT) calculations. NAP-XPS measurements reveal CO$_2$-like surface intermediates only under oxygen-rich conditions, consistent with oxygen-assisted CO oxidation. DFT calculations support an oxygen-assisted co-adsorption pathway in which CO interacts preferentially with adsorbed oxygen species stabilized near Zn-deficient sites, forming weakly bound CO$_2$-like structures. These results identify oxygen coverage, rather than defect density alone, as the key factor controlling CO$_2$-like intermediate formation on defective ZnS and establish defective ZnS as a model platform for studying oxygen-assisted surface chemistry on non-oxide semiconductors.
\end{abstract}

\ \ CO oxidation remains a cornerstone reaction in heterogeneous catalysis, serving both as a prototypical system for mechanistic investigations and as a key elementary step in emerging carbon management strategies \cite{freund2011co,kersell2020co}. In this reaction, CO species are oxidized to CO$_2$, which can subsequently participate in hydrogenation pathways leading to value-added chemicals and fuels \cite{van2017surface}.  From a materials science and surface chemistry perspective, catalytic performance is intimately linked to the nature of the catalyst and to the elementary steps governing the reaction mechanism. In this context, traditional catalysts, including noble metals such as Pd, Pt, and transition metal oxides, have been widely explored for the CO oxidation reaction \cite{wang2024recent,luneau2019dilute,shen2014study}. 

Recently, metal sulfide materials have emerged as an attractive alternative class of materials. Their oxygen-free lattice provides a well-defined platform to probe oxygen activation and reaction intermediates with reduced ambiguity, particularly when using surface-sensitive techniques such as near ambient pressure X-ray photoelectron spectroscopy (NAP-XPS) \cite{he2022recent, jurgensen2018situ, hedevang2026situ}. Yet, their catalytic activity is often limited due to the low density of active sites. Defect engineering, including the introduction of substitutional dopants and vacancies, has been shown to create new reactive centers and enhance catalytic performance by modifying adsorbate-catalyst interaction, the local electronic structure and adsorption properties \cite{muhammad2024defect,liang2021defect,zou2025boosting}.  Despite these advances, it is unclear whether defect-driven CO oxidation under reactive conditions proceeds independently of the surface CO:O$_2$  ratio ($\chi$) on sulfide materials.

Zinc sulfide is one of the most well-known materials in this class, with recent advances disclosing the role of defects on electronic, optical, and catalytic properties \cite{de2025formation,luo2023synthesis,pang2019cation}. While pristine ZnS displays a high resistivity and negligible active sites, the formation of zinc vacancies enhances the CO$_2$ reduction reaction, improves the surface conductivity, and increases the number of active sites such that excessive oxygen exposure results in the formation of a ZnO/ZnS interface \cite{dengo2018thermal}. These recent advances position defective ZnS surfaces as a promising model system to understand the mechanism governing CO oxidation on metal sulfide systems. Within Langmuir-Hinshelwood mechanism (LH), both reactants must adsorb on the catalyst surfaces, enabling chemisorbed CO to react with dissociated oxygen species to form CO$_{2}$ \cite{krick2016co} Alternatively, CO can react with weakly bound topmost oxygen of an oxygen-containing catalyst, forming CO$_{2}$ while leaving an oxygen vacancy behind, consistent with the Maars-van Krevelen mechanism (MvK) \cite{mars1954oxidations} A third possibility involves a direct reaction between gas-phase CO and lattice oxygen, following an Eley–Rideal (ER) pathway \cite{van2017surface}. For defect-rich sulfides, identifying the reaction pathway is critical, as it determines whether defects can intrinsically sustain catalytic turnover under equilibrium feeds ($\chi =1$) or whether an oxygen-rich environment ($\chi < \ 1$ )is required to enable CO$_2$ formation. 

  Herein, we investigate the CO oxidation on defective ZnS surfaces. By combining NAP-XPS and  DFT calculations, we reveal that CO molecules adsorb at oxygen sites through a co-adsorption mechanism. Notably, CO$_2$-like surface intermediates are detected only under an oxygen-rich scenario ($ \chi =0.5$), indicating that defects alone are insufficient to promote detectable CO activation under the present near-ambient-pressure conditions. For $ \chi = 1 $, the surface undergoes partial oxidation without the formation of CO$_2$ intermediates. DFT simulations reveal that, after anchoring on an oxygen site, the CO$_2$-like structure leaves the surface and weakly interacts with the ZnS surface, in a MvK-ER-like fashion. To the best of our knowledge, this work delivers the first pressure-dependent mechanistic description of CO oxidation on ZnS and defines the conditions under which defect-rich sulfides can engage surface-chemistry species under near-realistic environments.

\begin{figure}[h!]
    \centering
    \includegraphics[width=1\textwidth]{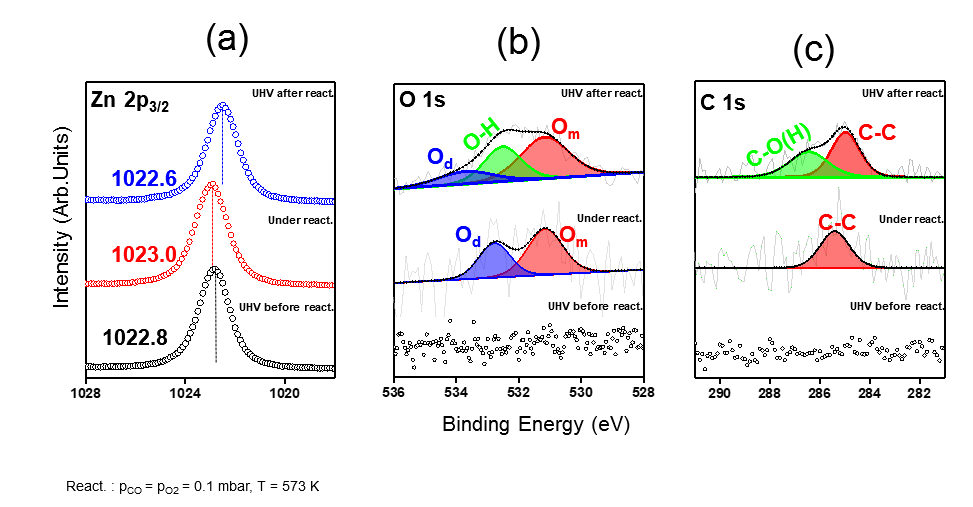}
    \caption{CO oxidation in the CO:O = 1:1 scenario. Zn 2p$_{3/2}$ (a), O 1s (b), and C 1s (c) were recorded in three different contexts: UHV before reaction (bottom panels), under reaction (middle panels), and UHV after reaction (top panels). Both UHV condition measurements were taken at room temperature, while operando conditions experiments were carried out at 0.1 mbar of CO and 0.1 mbar of O$_2$ at 573 K, for 60 min, as highlighted in the bottom-left corner. Light gray line,  black line-connected filled dots, and filled areas denote the raw, envelope, and components fitted data, respectively.}
    \label{stoic}
\end{figure}

It has been previously shown that catalytic activity on ZnS surfaces is boosted in the presence of native defects. In the context of oxidation reaction, $O_2$ readily dissociates ($O_2 \rightarrow 2 O_{ads}$), resulting in the partial oxidation of ZnS surfaces \cite{de2025growth}. Regarding $CO_2$ adsorption on ZnS,  while more active sites derived from Zn deficiency provide $CO_{2}^{\delta -}$ as intermediate, an oxygen-rich scenario gives rise to $CO_{3}^{-}$-like intermediates \cite{de2026insights}. Motivated by the role of defects on ZnS surfaces and recent insights into the interaction between $CO_{2}$ and ZnS, this work investigates the CO oxidation reaction under two distinct regimes: (i) $\chi = 1$: ($ p_{CO} = p_{O_{2}}$)  and (ii) $\chi= 0.5$ ($ p_{CO} =0.5 \  p_{O_{2}}$). For this purpose, a cleaned Zn-deficient ZnS (001) surface was simultaneously exposed to a CO ($p_{CO}$) and $O_2$ ($p_{O_{2}}$) atmosphere, and heated at 573 K during 60 minutes (See SI for experimental details regarding the cleaning cycles). NAP-XPS measurements in the scenario (i) were carried out by exposing the system to $p_{CO}$ = $p_{O_{2}}$ = $0.1 \ mbar$, revealing the fingerprints depicted in \textbf{Figure} \ref{stoic}. The Zn 2p line shape remains unchanged throughout the reaction, although its binding energy shifts by circa 0.2 eV (\textbf{Figure} \ref{stoic} (a) top panel). Binding energy deviations under operando conditions are commonly attributed to the interaction between photoelectrons and gas phase molecules in the chamber \cite{teschner2024understanding}. However, in the context of defective ZnS surfaces, a low binding energy shift is rather related to partial oxidation \cite{de2025growth,de2026insights}. This interpretation correlates well with the features observed in the O 1s region, as depicted in \textbf{Figure} \ref{stoic} (b). The cleaned ZnS surface initially shows no oxygen contribution (bottom panel). Upon CO + O$_2$ dosing (middle panel), the dissociation of oxygen molecules occurs ($O_2 \rightarrow 2O_{ads}$),  resulting in two main components: One at 531 eV, related to oxygen near metallic sites ($O_m$), and an additional feature at 532.7 eV, attributed to the interaction between oxygen-containing adsorbate and the partially oxidized catalyst surface ($O_d$) \cite{frankcombe2023interpretation}. The O$_m$ and O$_d$ assignments are based on their binding energies, their appearance only after O$_2$ exposure, and their correlation with the Zn 2p$_{3/2}$ shift. The fitting constraints, full-width-at-half-maximum values, and relative areas are provided in table S1 and \textbf{Figure} S2.
After returning to UHV at room temperature (\textbf{Figure} \ref{stoic} (b) top panel), the O~1s envelope is well described by including an additional component attributed to OH species, likely derived from the desorption of residual water from the chamber, which can also contribute to a small deviation of the $O_d$ binding energy ($\Delta E(O_{d }) = 0.23 \ eV$). If $O_d$ were primarily associated with CO oxidation intermediates, a correlated C~1s contribution in the 288–289.5 eV region would be expected. Yet, C 1s envelope analysis does not support this result, as depicted in \textbf{Figure} \ref{stoic} (c). Apart from the cleaned surface (\textbf{Figure} \ref{stoic} (c) bottom panel), whose featureless carbon region confirms the absence of initial carbon contamination, the surface under operando conditions (middle panel) reveals a single C–C component centered at 285 eV \cite{patel2025initiation}. The absence of an additional contribution in the 288–289.5 eV range—typically attributed to a $CO_{2}$ fingerprint—indicates that no CO$_2$ surface intermediate is detected within the sensitivity of our NAP-XPS measurements. Although the formation of transient $CO_{2}$ gas phase that readily desorbs from the surface cannot be ruled out, it seems that oxygen activation in the 1:1 regime is not sufficient to provide active sites that enable CO species to undergo oxidation.  After evacuation (top panel), the C 1s envelope displays an additional component related to C–O(H) species, consistent with the hydroxyl contributions identified in the O 1s region.


\begin{figure} [h!]
    \centering
    \includegraphics[width=1\textwidth]{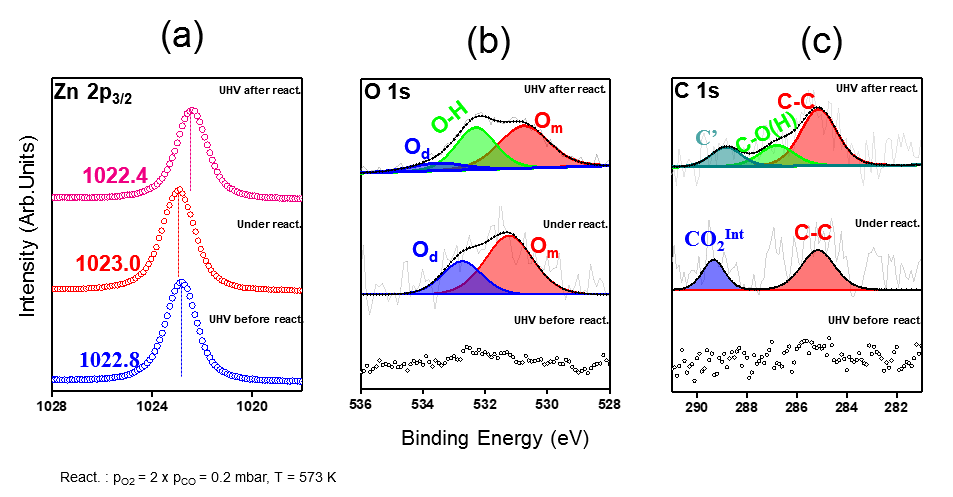}
    \caption{CO oxidation in the CO:O = 1:2 scenario. Zn 2p$_{3/2}$ (a), O 1s (b), and C 1s (c) signature evolution throughout all reaction scenarios, namely UHV before Reaction (bottom), under reaction (middle), and after reaction (top). UHV condition measurements were taken at room temperature, while Operando conditions experiments were carried out under 0.1 mbar of CO and 0.2 mbar of oxygen at 573 K, for 60 min. Light gray line,  black line-connected filled dots, and filled areas denote the raw, envelope, and components fitted data, respectively. Although minor intensity fluctuations are present in the operando spectra, introducing additional components would require unphysically narrow FWHM values and would not improve the fit in a chemically meaningful manner. The reader can see the fitting procedure and the raw data in the SI. }
    \label{Orich}
\end{figure}

        After a short UHV annealing at 973 K, the sample was exposed to an oxygen-rich atmosphere. In this scenario, the surface behavior toward CO oxidation markedly differs from that observed in case (i), as depicted in \textbf{Figure} \ref{Orich}. Although the Zn~2p core level spectra exhibit the same qualitative binding-energy shift previously observed, the magnitude of the shift is slightly larger, as shown in \textbf{Figure} \ref{Orich} (a), indicating a higher degree of surface oxidation. Consistently, the O 1s spectra in \textbf{Figure} \ref{Orich}(b) (middle panel) reveal an enhanced contribution from both the $O_m$ and $O_d$ components under reaction conditions. Upon evacuation, an OH-related feature emerges at approximately 532~eV, accompanied by a decrease in the absolute intensity of the $O_d$ component. The features in the C 1s envelope shown in \textbf{Figure} \ref{Orich} (c) emphasize the key results. The system under UHV before the reaction displaying no traces of carbon (\textbf{Figure} \ref{Orich} (b) bottom panel), changes under reaction by revealing a new component at 289 eV $CO_{2}^{int}$, in addition to the C-C signature at 285 eV. This new feature is attributed to intermediate CO$_2$ contributions, indicating formation of CO$_2$-like or carbonate-like surface species under oxygen-rich conditions. After evacuation, the intermediate features together with C–O(H) species derived from the desorption of water molecules likely recombine, resulting in the $C'$ component, as revealed in \textbf{Figure} \ref{Orich} (c) top panel.  Interestingly, from the 1:1 to the 1:2 scenario, the $O_d$ relative concentration under reaction drops from 40.65 \%  to 31.22 \%  (see SI). Although this observation alone does not unambiguously identify the active site, the correlation with the emergence of CO$_2$-like features in the C 1s region suggests a mechanism in which oxygen-containing surface species participate directly in CO adsorption and activation. Based on these experimental observations, we propose the following reaction pathway: oxygen molecules first dissociatively adsorb close to defect-related sites on ZnS surfaces. In the dynamic picture, further oxygen species may continuously dissociate either on metallic sites or adsorbed oxygen sites, explaining the $O_d$ and $O_m$ features observed in both scenarios. For $\chi = 1$, the steady-state concentration of O$_{ads}$ remains insufficient to stabilize detectable CO$_2$-like intermediates. In contrast, an oxygen-rich environment increases the surface concentration of reactive oxygen species, enabling CO species to anchor onto these sites and form CO$_2$-like intermediates, via a co-adsorption mechanism.

\begin{figure}[h!]
    \centering
    \includegraphics[width=1\textwidth]{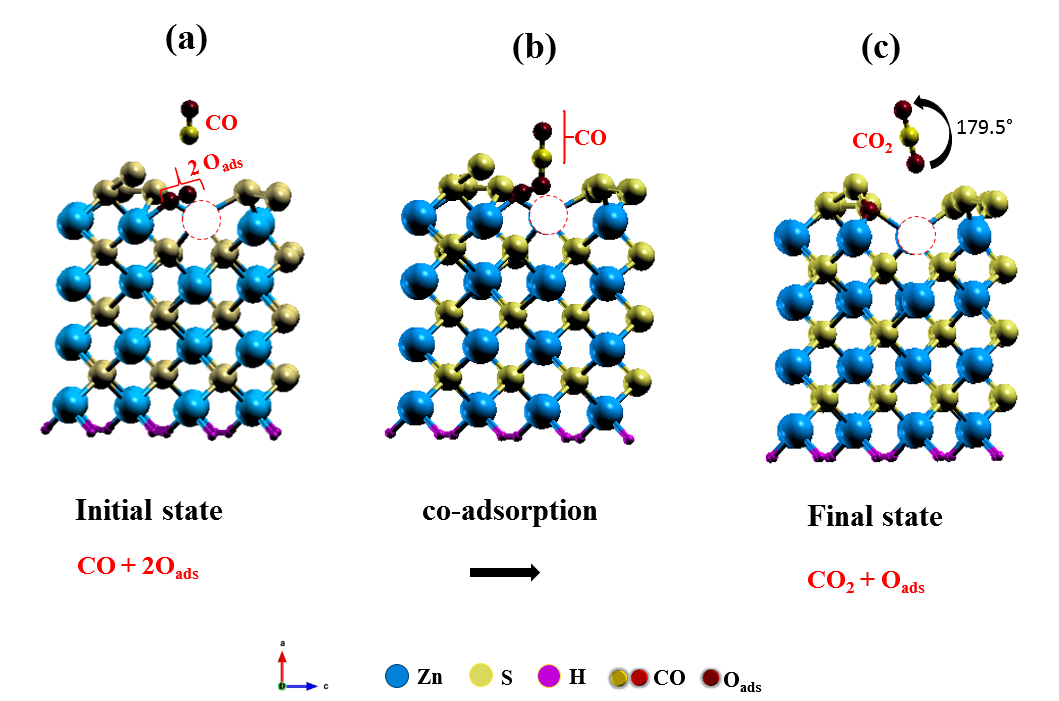}
    \caption{Side view of the DFT simulated relaxation pathway of a CO molecule interacting with an oxygen-adsorbed in a defective ZnS (001) surface. (a) Initially, CO in the gas phase will interact with a ZnS surface featuring two adsorbed oxygen atoms, derived from the dissociative adsorption of oxygen species. (b) Next, the CO molecule adsorbs atop one oxygen site, resulting in a CO$_2$-like structure. (c) Finally, the intermediate CO$_2$ desorbs from the surface after reaching its energetically favorable configuration. The red dotted circle denotes a Zn vacancy site.}
    \label{DFT}
\end{figure}

The CO oxidation process under oxygen-rich conditions was investigated using density functional theory calculations, as illustrated in \textbf{Figure}~\ref{DFT}. Given the experimental findings, we simulated the evolution of a CO molecule interacting with a defective ZnS slab ($O_{ads}/ZnS)$ featuring a zinc vacancy, highlighted by dotted circles in \textbf{Figure}~\ref{DFT}, and two adsorbed oxygen atoms resulted from the dissociation of oxygen molecules. The stability of oxygen species close to vacancy-related sites was investigated in a previous work, where we demonstrated that oxygen adsorption at the Zn-deficient ZnS surface is a highly exothermic process \cite{de2026insights}. 
In this scenario, the CO anchors onto an oxygen site rather than directly to Zn or S lattice sites, as depicted in \textbf{Figure}~\ref{DFT}(b). 
This co-adsorbed intermediate state seems to be the critical point for activating CO activation on ZnS. As recently reported in the literature, the CO adsorption on non-defective ZnS monolayer is very weak \cite{chhana2022theoretical}. In our case, even in the presence of zinc defects, the CO barely interacts with ZnS surfaces (see SI). This finding explains why the MvK-LH mechanism is not energetically favorable, since it requires CO adsorption as a preliminary step. Further relaxation leads the system to evolve toward a configuration in which CO$_2$ like species desorbs from the surface, while one oxygen atom remains stabilized near the vacancy center, as shown in \textbf{Figure}~\ref{DFT}(c). The O-O separation in this scenario closely resembles that of a molecular CO$_2$ species, providing direct theoretical evidence for the formation of CO$_2$ via a co-adsorption mechanism. The reaction global minimum then corresponds to a weakly bound, nearly linear CO$_2$ molecule with a bond angle of approximately 179.5$^\circ$, with an exothermic adsorption energy ($E_{ads} = -0.21 \ eV$.  The slight deviation from linearity suggests residual interaction with the surface and partial charge transfer, consistent with the experimentally observed CO$_2^{\delta-}$ fingerprint. 

 \begin{figure}[h!]
    \centering
    \includegraphics[width=1.0\textwidth]{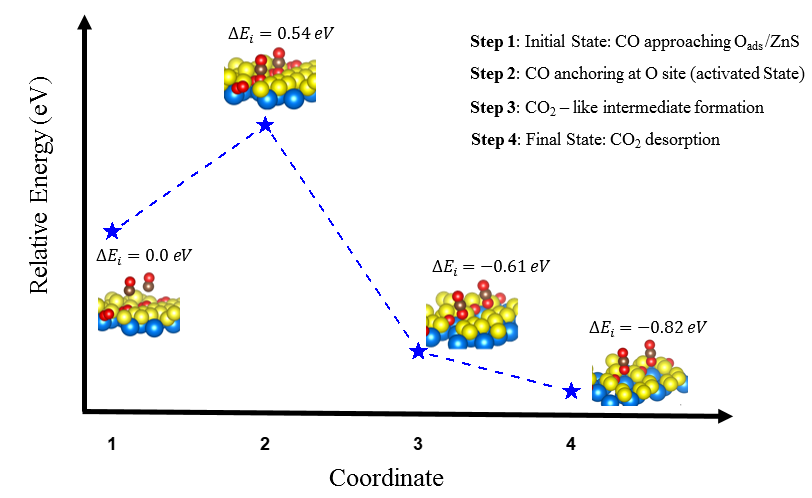}
    \caption{Geometric-guided DFT reaction pathway: From the CO gas phase interacting with adsorbed oxygen defective ZnS surface (Initial State) to the CO$_2$-like product derived from CO-co-adsorption reaction.}
    \label{DFT-energetic}
\end{figure}

Insights into the energetic pathway are derived by computing the configuration-dependent interaction energy $E_{i}$ at each reaction step \cite{kersell2020co}, as depicted in Figure \ref{DFT-energetic}. In our model, the CO oxidation is described by the following reaction: $ CO_{(g)} + 2O_{ads} \rightarrow CO_{2} + O_{ads}$. In this framework, the initial configuration, corresponding to a CO molecule approaching the O$_{ads}$/ZnS surface, is taken as the reference ($\Delta E_{i} = 0 \ eV)$. The interaction energy at each step is defined by $E_{i} = E_{all} - E_{free} - E_{mol}$, where $E_{all}$ is the total energy of the interacting system (adsorbate + O/ZnS), $E_{free}$ corresponds to the system without adsorbate, and $E_{mol}$ is the total energy of the individual molecule evaluated in the same geometry as in the full configuration (see SI). The highest energy barrier corresponds to the CO adsorption onto an oxygen site, with $\Delta E_{i} = 0.544 \ eV $. This configuration exhibits a carbonate-like structure, in good agreement with NAP-XPS experiments. Following this step, the system undergoes a pronounced stabilization as the structure evolves into a weakly bound CO$_{2}$-like intermediate (Step 3), resulting in an oxidized ZnS surface interacting with a CO$_2$-like adsorbate.  This weak interaction is highlighted by the lower energy barrier to reach the final state. ($\Delta E = -0.21 \ eV $). This result is consistent with the calculated adsorption energy of a CO$_{2}$-like structure interacting with an oxidized defective ZnS slab ($E_{ads}^{CO_{2}-like} = -0.16 \ eV $). Although representing an exothermic reaction,  this adsorption energy is lower compared to other metal catalyst surfaces, in good agreement with our previous investigation on CO$_{2}$ adsorption on defective ZnS \cite{de2026insights}. Overall, the calculated reaction pathway indicates that CO adsorption is significantly facilitated in the presence of oxygen-stabilized defective ZnS surfaces. After O$_2$ dissociation, one of the resulting $O{_\text{ads}}$ species is weakly bound to the ZnS surface near a vacancy center, lowering the barrier for CO to anchor and form CO$_2$ via a co-adsorption configuration. Importantly, the oxygen coverage is critical. When only a single $O_{ads}$ is present near a vacancy center on the surface, the activation barrier becomes significantly higher (3.53 eV, as shown in SI), which can be attributed to its stronger binding to ZnS, increasing the energetic cost required for rearrangement and CO accommodation. Consistently, the final-state interaction energy of 0.43 eV (see SI) indicates that CO$_2$ formation through co-adsorption at low $O_{ads}$ coverage is energetically unfavorable. As such, the higher the oxygen concentration in the ZnS lattice, the higher the probability of a CO molecule meeting one oxygen site. Kinetically, this indicates that increasing the $O_{ads}$ concentration decreases the time for a CO molecule to meet one oxygen site, improving the reaction stability. Although our calculations cannot capture the time scale in which the co-adsorption takes place, they explain the microscopic mechanism behind the experimental finding that suggests CO oxidation occurs only under oxygen-rich conditions ($\chi < 1 $), proceeding in a MvK-ER fashion.

By combining NAP-XPS and DFT calculations, we were able to obtain insights into the mechanism of CO oxidation on defective ZnS surfaces. While defects are important to increase the number of active sites that could favor the reaction, the balance between CO and O$_2$ is the central point for promoting the reaction. In particular, the reaction proceeds only under oxygen-rich conditions, emphasizing that high surface oxygen coverage is required to drive the process. For $\chi = 1$, the oxidation of the ZnS takes place rather than the expected CO oxidation, as confirmed by NAP-XPS data revealing an oxygen bound to metal without CO$_{2}$ in the C 1s envelope. In an oxygen-rich scenario ($\chi = 2 $), the targeted reaction takes place. The C 1s spectrum under reaction reveals an additional component at 288.8 eV, suggesting the partial formation of $CO_{2}$ intermediates. According to DFT calculations, CO molecules preferentially anchor on oxygen sites, giving rise to $CO_{2}$-like species via a co-adsorption mechanism.

\section*{Supporting Information}
Experimental description, theoretical methods, fitting procedure, O 1s components relative concentration in the two scenarios explored, all experimental raw data, and the coordination of all structures used for DFT calculations (PDF).

\begin{acknowledgement}
The authors acknowledge the support from the funding agencies CNPq and FAPERJ. PRA and CC thank CAPES for support through the PIPD scholarship. MGM also acknowledges the support from INCT on Materials Informatics. PRA, PV and MGM acknowledge the support of CENAPAD and LNCC for the computational resources employed in this work.
\end{acknowledgement}

\bibliography{Reference}

\end{document}